\documentclass[preprint,showpacs,preprintnumbers,amsmath,amssymb]{revtex4}

\pdfoutput=1

\usepackage{graphicx}
\usepackage{dcolumn}
\usepackage{bm}
\usepackage{amssymb}

\begin{document}
\title{
Dynamical properties of an exactly solvable coupled quantum double-well system:
The evolution speed and entanglement
}

\author{Hideo Hasegawa}
\altaffiliation{hideohasegawa@goo.jp}
\affiliation{Department of Physics, Tokyo Gakugei University,  
Koganei, Tokyo 184-8501, Japan}%

\date{\today}

\begin{abstract}
We have studied dynamical properties of 
an exactly solvable quantum coupled double-well (DW) system 
with Razavy's hyperbolic potential.
With the use of four kinds of initial wavepackets, 
the correlation function $\Gamma(t)$ and the concurrence $C(t)$
which is a typical measure of the entanglement in two qubits, are calculated.
We obtain the orthogonality time $\tau$ which signifies 
a time interval for an initial state to evolve to its orthogonal state,
and the temporal average of $C_{av}$ $(=\sqrt{\langle C(t)^2 \rangle})$.
The coupling dependence of $\tau$ and the concurrence [$C_{av}$ or $C(0)$], and 
the relation between $\tau$ and the concurrence are investigated.
Our calculations have shown that the evolution speed measured by $\tau^{-1}$ 
is not necessarily increased with increasing the concurrence in coupled DW systems.

\vspace{0.5cm}

\end{abstract}

        

\maketitle
\newpage
\section{Introduction}
The two-level (TL) system 
has been employed for a study on qubits which play important roles 
in quantum information and quantum computation \cite{Ref1}.
The connection between the quantum evolution speed and the entanglement 
has been extensively studied with the use of the TL model
\cite{Margolus98,Pfeifer93,Giovannetti03,Giovannetti03b,Batle05,Curilef06,Borras06,Chau10,Zander13}.
It has been pointed out that the speed of evolution in certain quantum state may be measured by
the orthogonality time which expresses a time for an initial state
to reach its orthogonal state \cite{Margolus98,Pfeifer93,Giovannetti03,Giovannetti03b}.
Margolus and Levitin \cite{Margolus98} asserted that the orthogonal time $\tau$
is given by $\tau \geq \pi \hbar/(2E)$ where $E$ stands for the expectation energy
of a given quantum system relative to the ground-state energy. 
This result complements the time-energy uncertainty relation 
requiring $\tau \geq \pi \hbar/(2 \:\Delta E)$ where $\Delta E$ expresses
the root-mean-square value of the system energy \cite{Pfeifer93}.
Combining the above two results \cite{Margolus98,Pfeifer93}, 
Giovannetti {\it et al.} \cite{Giovannetti03,Giovannetti03b} pointed out 
that the entanglement permits to achieve the maximum evolution speed measured 
by $\tau_{\min}$ which is given by
\begin{eqnarray}
\tau &\geq& \tau_{\min} 
\equiv 
\max \left( \frac{\pi \hbar}{2 E}, \;\;\frac{\pi \hbar}{2 \:\Delta E} \right).
\label{eq:J1}
\end{eqnarray}
Batle {\it et al.} \cite{Batle05} and Curilef {\it et al.} \cite{Curilef06}
showed that in two uncoupled qubits,
the ratio of $\tau/\tau_{min}$ is unity for a maximally entangled state
and $\sqrt{2}$ for a separate state \cite{Batle05,Curilef06}.
Borr\'{a}s {\it et al.} \cite{Borras06} made an extension of Ref. \cite{Batle05}
for two uncoupled qubits,
showing a clear correlation between the evolution speed and concurrence.
It was pointed out by Chau \cite{Chau10} that for the singular case with $\vert a_3 \vert^2=0$
which was not discussed in Refs. \cite{Borras06,Batle05}, the relation between
entanglement and $\tau$ can be very different from the generic case 
with $\vert a_3 \vert^2 \neq 0$, where $a_3$ means the expansion coefficient in
a wavepacket [Eq. (\ref{eq:G1})].
A concept of the orthogonality time is generalized to the case where an initial state
evolves to an arbitrary final state \cite{Giovannetti03b,Borras06}.
Effects of interactions between two qubits which modify the entanglement
are investigated in Refs. \cite{Giovannetti03,Zander13}.
Zander {\it et al.} \cite{Zander13} have made a detailed study on the relation 
between the ratio of $\tau/\tau_{min}$ and the entanglement in interacting two qubits.
It is shown that, with the exception of some marginal special cases, only initial states 
with low entanglement tend to evolve in the fastest way in coupled qubits \cite{Zander13}.
Related discussion will be given in Sec. IV. 

Double-well (DW) potential models have been widely employed in various fields
of quantum physics. 
Although quartic DW potentials are commonly adopted for the theoretical study,
one cannot obtain their exact eigenvalues and eigenfunctions
of the Schr\"{o}dinger equation.
Then it is necessary to apply various approximate approaches
such as perturbation and spectral methods to quartic potential models \cite{Tannor07}.
Razavy \cite{Razavy80} proposed the quasi-exactly solvable hyperbolic DW potential, 
for which one may exactly determine a part of whole eigenvalues and eigenfunctions.
A family of quasi-exactly solvable potentials has been investigated
\cite{Finkel99,Bagchi03}.
In contrast to the TL model which is a simplified model of a DW system, studies 
on {\it coupled} DW systems are scanty, as far as we are aware of.
This is because a calculation of a coupled DW system is much tedious than 
those of a single DW system and of a coupled TL model. 
In the present study, we adopt coupled two DW systems, each of which is described by Razavy's potential. 
One of advantages of our model is that we may exactly determine eigenvalues 
and eigenfunctions of the coupled DW system. We study dynamics of wavepackets,
calculating the correlation function $\Gamma(t)$ by which the orthogonality time $\tau$
is obtained, and the concurrence $C(t)$
which is one of typical measures of entanglement.
We investigate the relation between the speed of quantum evolution measured by $\tau^{-1}$ and 
the entanglement expressed by the concurrence. 
The difference and similarity between results in our coupled DW system and 
the TL model \cite{Giovannetti03,Giovannetti03b,Batle05,Curilef06,Borras06,Zander13} are discussed.
These are purposes of the present paper.

The paper is organized as follows.
In Sec. II, we describe the calculation method employed in our study,
briefly explaining Razavy's potential \cite{Razavy80}.
Exact analytic expressions for eigenvalues and eigenfunctions for
coupled DW systems are presented. In Sec. III,
with the use of four kinds of initial wavepackets, we perform model calculations
of the time-dependent correlation $\Gamma(t)$ and concurrence $C(t)$, 
evaluating the orthogonality time $\tau$ and 
temporal average of concurrence $C_{av}$ $(= \sqrt{\langle C(t)^2 \rangle})$.
The relation between the calculated $\tau$ and the concurrence, $C_{av}$ or $C(0)$, 
is studied. Sec. IV is devoted to our conclusion.

\section{The adopted method}
\subsection{Coupled double-well system with Razavy's potential}
We consider coupled two DW systems whose Hamiltonian is given by 
\begin{eqnarray}
H &=& \sum_{n=1}^2 \left[ -\frac{\hbar^2}{2m} \frac{\partial^2}{\partial x_n^2} 
+ V(x_n )\right] - g x_1 x_2, 
%
\label{eq:H1}
\end{eqnarray}
with
\begin{eqnarray}
V(x) &=& \frac{\hbar^2 \kappa^2}{2m}
\left[\frac{\xi^2}{8} \:{\rm cosh} \:4 \kappa x - 4 \xi \:{\rm cosh} \:2 \kappa x- \frac{\xi^2}{8}
\right],
\label{eq:H3}
\end{eqnarray}
where $x_1$ and $x_2$ stand for coordinates of two distinguishable particles 
of mass $m$ in double-well systems coupled by an interaction $g$, and
Razavy's potential $V(x)$ depends on two parameters of $\xi$ and $\kappa$ \cite{Razavy80}.
The potential $V(x)$ with $\hbar=m=\xi=\kappa=1.0$ adopted in this study
is plotted in Fig. 1(a).
Minima of $V(x)$ locate at $x_s=\pm 1.38433$ with $V(x_s)=-8.125$
and its maximum is $V(0)=-2.0$ at $x=0.0$ \cite{Note2}.

First we consider the case of $g=0.0$ in Eqs. (\ref{eq:H1}) and (\ref{eq:H3}).
Eigenvalues of Razavy's double-well potential of Eq. (\ref{eq:H3}) are given by \cite{Razavy80}
\begin{eqnarray}
\epsilon_0 &=& \frac{1}{2}\left[ -\xi -5 -2 \sqrt{4-2 \xi+\xi^2} \right], \\
\epsilon_1 &=& \frac{1}{2}\left[ \xi-5 -2 \sqrt{4+2 \xi+\xi^2} \right], \\
\epsilon_2 &=& \frac{1}{2}\left[ -\xi-5 +2 \sqrt{4-2 \xi+\xi^2} \right], \\
\epsilon_3 &=& \frac{1}{2}\left[ \xi-5 +2 \sqrt{4+2 \xi+\xi^2} \right]. 
\end{eqnarray}
Eigenvalues for the adopted parameters are $\epsilon_0=-4.73205$, $\epsilon_1=-4.64575$,
$\epsilon_2=-1.26795$ and  $\epsilon_3=0.645751$.
Both $\epsilon_0$ and $\epsilon_1$ locate below $V(0)$ as shown by dashed curves in Fig. 1(a),
and $\epsilon_2$ and $\epsilon_3$ are far above $\epsilon_1$. In this study,
we take into account the lowest two states of $\epsilon_0$ and $\epsilon_1$ 
whose eigenfunctions are given by \cite{Razavy80}
\begin{eqnarray}
\phi_0(x) &=& A_0 \; e^{-\xi \:{\rm cosh} \:2x/4} \left[3 \xi \:{\rm cosh} \:x
+(4-\xi+2 \sqrt{4-2 \xi+\xi^2})\: {\rm cosh}\: 3x \right], \\
\phi_1(x) &=&  A_1 \;e^{-\xi \:{\rm cosh} \:2x/4} \left[3 \xi \:{\rm sinh}\: x
+(4+\xi+2 \sqrt{4+2 \xi+\xi^2})\: {\rm sinh} \:3x \right],
\end{eqnarray}
$A_{n}$ ($n=0,1$) denoting normalization factors.
Figure 1(b) shows the eigenfunctions of $\phi_0(x)$ and $\phi_1(x)$, which 
are symmetric and anti-symmetric, respectively, with respect to the origin.

\begin{figure}
\begin{center}
\includegraphics[keepaspectratio=true,width=120mm]{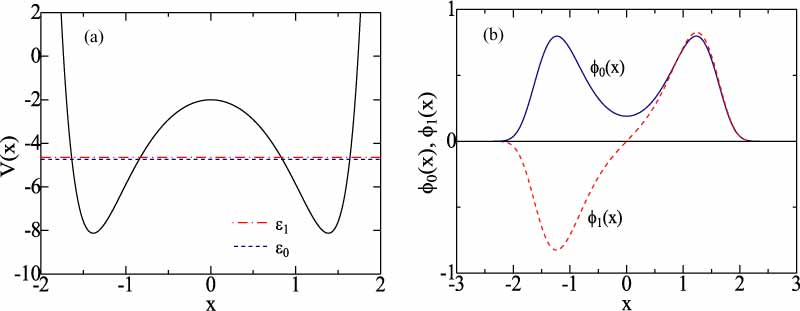}
\end{center}
\caption{
(Color online) 
(a) Razavy's DW potential $V(x)$ (solid curve),
dashed and chain curves expressing eigenvalues of $\epsilon_0$ and $\epsilon_1$,
respectively, for $\hbar=m=\xi=\kappa=1.0$ [Eq.(\ref{eq:H3})].
(b) Eigenfunctions of $\phi_0(x)$ (solid curve) and $\phi_1(x)$ (dashed curve).
}
\label{fig1}
\end{figure}

\subsection{Eigenvalues and eigenstates of the coupled DW system}
We calculate exact eigenvalues and eigenstates of the coupled two DW systems described 
by Eq. (\ref{eq:H1}). With basis states of 
$\phi_0 \phi_0$, $\phi_0 \phi_1$, $\phi_1 \phi_0$ and $\phi_1 \phi_1$
where $ \phi_n \phi_k \equiv  \phi_n(x_1) \phi_k(x_2)$,
the energy matrix for the Hamiltonian given by Eq. (\ref{eq:H1}) is expressed by
\begin{eqnarray}
{\cal H} &=& \left( {\begin{array}{*{20}c}
   {2 \epsilon_0 } & {0 } & {0 } & {-g \gamma^2} \\
   {0 } & {\epsilon_0 + \epsilon_1 } & {-g \gamma^2 } & {0} \\
   {0 } & {-g \gamma^2 } & {\epsilon_0 + \epsilon_1 } & {0} \\
   {-g \gamma^2 } & {0 } & {0 } & {2 \epsilon_1} \\   
\end{array}} \right),
\label{eq:H4}
\end{eqnarray}
with
\begin{eqnarray}
\gamma &=& \int_{-\infty}^{\infty} \phi_0(x)\: x \: \phi_1(x)\:dx=1.13823.
\label{eq:H5}
\end{eqnarray}
Eigenvalues of the energy matrix are given by
\begin{eqnarray}
E_0 &=& \epsilon -\sqrt{\delta^2+ g^2 \gamma^4}, 
\label{eq:H6a}\\
E_1 &=& \epsilon - g \gamma^2, \\
E_2 &=& \epsilon + g \gamma^2, \\
E_3 &=& \epsilon + \sqrt{\delta^2+ g^2 \gamma^4},
\label{eq:H6b}
\end{eqnarray}
where
\begin{eqnarray}
\epsilon &=& \epsilon_1+\epsilon_0=-9.3778, \\
\delta &=& \epsilon_1-\epsilon_0=0.0863.
\end{eqnarray}
Corresponding eigenfunctions are given by
\begin{eqnarray}
\Phi_0(x_1,x_2) &=& \cos \theta \:\phi_0(x_1) \phi_0(x_2) 
+ \sin \theta \:\phi_1(x_1) \phi_1(x_2), 
\label{eq:H7a}\\
\Phi_1(x_1,x_2) &=& \frac{1}{\sqrt{2}} \left[ \phi_0(x_1) \phi_1(x_2)
+ \phi_1(x_1) \phi_0(x_2) \right], \\
\Phi_2(x_1,x_2) &=& \frac{1}{\sqrt{2}} \left[- \phi_0(x_1) \phi_1(x_2)
+ \phi_1(x_1) \phi_0(x_2) \right], \\
\Phi_3(x_1,x_2) &=& -\sin \theta \:\phi_0(x_1) \phi_0(x_2)
+ \cos \theta \: \phi_1(x_1) \phi_1(x_2),
\label{eq:H7b}
\end{eqnarray}
where
\begin{eqnarray}
\tan \:2 \theta &=& \frac{g \gamma^2}{\delta}.
\;\;\;\;\mbox{$\left(-\frac{\pi}{4} \leq \theta \leq \frac{\pi}{4} \right)$}
\label{eq:H7c}
\end{eqnarray}

Eigenvalues $E_{\nu}$ ($\nu=0-3$) are plotted as a function of $g$ in Fig. 2, which is
symmetric with respect to $g=0.0$. 
For $g=0.0$, $E_1$ and $E_2$ are degenerate. We hereafter study the case of $g \geq 0.0$.
With increasing $g$, energy gaps between $E_0$ and $E_1$ and between $E_2$ 
and $E_3$ are gradually decreased while that between $E_1$ and $E_2$ is increased.
We note that differences between eigenvalues 
defined by $\Omega_{\nu}=(E_{\nu}-E_0)/\hbar$
satisfy the relation:
\begin{eqnarray}
\Omega_3 &=& \Omega_1+\Omega_2.
\label{eq:H8}
\end{eqnarray}
For $g=0.0$, we obtain $\Omega_1=\Omega_2=\Omega_3/2$.

\begin{figure}
\begin{center}
\includegraphics[keepaspectratio=true,width=70mm]{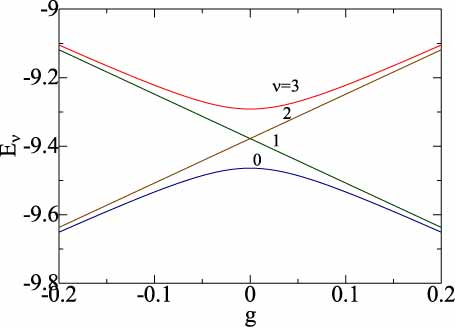}
\end{center}
\caption{
(Color online) 
Eigenvalues $E_{\nu}$ ($\nu=0-$3) of a coupled DW system as a function 
the coupling strength $g$.
}
\label{fig2}
\end{figure}

\subsection{The correlation function and orthogonality time}
The time-dependent wavepacket is expressed by
\begin{eqnarray}
\Psi(t) &=& \Psi(x_1,x_2,t)
= \sum_{\nu=0}^{3}\:a_{\nu} \: \Phi_{\nu}(x_1,x_2) \:e^{-i E_{\nu} t/\hbar},
\label{eq:G1}
\end{eqnarray}
where expansion coefficients $a_{\nu}$ satisfy the relation
\begin{eqnarray}
\sum_{\nu=0}^3 \vert a_{\nu} \vert^2 &=& 1.
\label{eq:G3}
\end{eqnarray}
The correlation function $\Gamma(t)$ is defined by
\begin{eqnarray}
\Gamma(t) &=& \vert \int_{-\infty}^{\infty} \int_{-\infty}^{\infty} 
\Psi^*(x_1,x_2, 0) \:\Psi(x_1,x_2,t)\;dx_1\:dx_2 \: \vert, \\
&=& \vert \; \vert a_0 \vert^2  + \sum_{\nu=1}^3 \: \vert a_{\nu} \vert^2 
\:e^{- i \Omega_{\nu} t} \: \vert.
\label{eq:G2}
\end{eqnarray}
The orthogonality time $\tau$ is 
provided by a time interval such that an initial wavepacket takes to evolve into
the orthogonal state \cite{Giovannetti03,Giovannetti03b,Batle05,Curilef06},
\begin{eqnarray}
\tau &=& \min_{\forall \:t \:> 0}\; \{ \Gamma(t)=0 \}.
\label{eq:E2}
\end{eqnarray}

In the case of wavepackets including only two states with
$a_{\nu}=(1/\sqrt{2}) \:(\delta_{\nu, 0}+\delta_{\nu, \kappa})$, 
the correlation function becomes
\begin{eqnarray}
\Gamma(t) &=& \frac{1}{2} \vert 1+ e^{-i \Omega_{\kappa} t} \vert
= \sqrt{ \frac{1+ \cos \Omega_{\kappa} t}{2} },
\label{eq:E3}
\end{eqnarray}
for which we easily obtain $\tau$
\begin{eqnarray}
\tau &=& \frac{\pi}{\Omega_{\kappa}}.
\label{eq:E4}
\end{eqnarray}
In the case of $g=0.0$, 
Eq. (\ref{eq:E2}) becomes
\begin{eqnarray}
\tau &=& \min_{\forall \:t}\; \large\{\vert a_0 \vert^2 
+ (\vert a_1 \vert^2+ \vert a_2 \vert^2)\:z(t) + \vert a_3 \vert ^2 z(t)^2 =0 \large\}, 
\end{eqnarray}
where $z(t)=e^{- i \Omega_1 t}$.
Solutions of $\tau$ may be obtainable from roots of respective 
polynomial equations for $z(t)$ \cite{Batle05,Borras06,Curilef06}.
In a general case, however, $\tau$ is obtainable 
by solving Eq. (\ref{eq:E2}) with a numerical method.

\subsection{The concurrence}
We have calculated the concurrence of coupled DW systems.
Substituting Eqs. (\ref{eq:H7a})-(\ref{eq:H7b}) into Eq. (\ref{eq:G1}), 
we obtain
\begin{eqnarray}
\vert \Psi \rangle &=& c_{00} \vert 0\;0 \rangle+ c_{01} \vert 0\;1 \rangle
+ c_{10} \vert 1\;0 \rangle + c_{11} \vert 1\;1 \rangle,
\label{eq:K1}
\end{eqnarray}
with
\begin{eqnarray}
c_{00} &=& a_0 \:\cos \theta \:e^{- i E_0 t}-a_3 \:\sin \theta \: e^{-i E_3 t}, 
\label{eq:K2a} \\
c_{01} &=&  \frac{1}{\sqrt{2}}(a_1 \:e^{- i E_1 t}-a_2 \:e^{- i E_2 t}), 
\label{eq:K2b} \\
c_{10} &=& \frac{1}{\sqrt{2}}(a_1\:e^{- i E_1 t}+a_2\:e^{- i E_2 t}),
\label{eq:K2c} \\
c_{11} &=& a_0\:\sin \theta\:e^{- i E_0 t}+a_3 \:\cos \theta\:e^{- i E_3 t},
\label{eq:K2d}
\end{eqnarray}
where $\vert k \;\ell \rangle=\phi_k(x_1) \phi_{\ell}(x_2)$ with $k, \ell=0,1$.
The concurrence $C$ of the state $\vert \Psi \rangle$ given by Eq. (\ref{eq:K1}) is defined by
\cite{Wootters01}
\begin{eqnarray}
C^2 &=& 4 \:\vert c_{00} c_{11}- c_{01} c_{10} \vert^2. 
\label{eq:K4}
\end{eqnarray}
The state given by Eq. (\ref{eq:K1}) becomes factorizable if and only if the
relation: $c_{00} c_{11}- c_{01} c_{10} =0$ holds.
Substituting Eqs. (\ref{eq:K2a})-(\ref{eq:K2d}) into Eq. (\ref{eq:K4}), 
we obtain the concurrence 
\begin{eqnarray}
C(t)^2 &=& \vert (a_0^2 - a_3^2 \:e^{-2i \Omega_3 t}) \sin 2 \theta
+ 2 a_0 a_3 \:\cos 2\theta \:e^{-i \Omega_3 t}
- a_1^2 \:e^{-2i \Omega_1 t}+ a_2^2 \:e^{-2 i \Omega_2 t} \vert^2,
\nonumber \\
&&
\label{eq:K4b}
\end{eqnarray}
whose initial value becomes
\begin{eqnarray}
C(0)^2 &=& \vert (a_0^2- a_3^2) \: \sin 2 \theta
+ 2 a_0 a_3 \:\cos 2\theta- a_1^2 + a_2^2 \vert^2.
\end{eqnarray}
We should note that the concurrence becomes time dependent in general for $g \neq 0.0$ 
because the coupling modifies the entanglement in two qubits, although
it is time-independent for uncoupling case ($g=0.0$) where $\theta=0.0$ and
$\Omega_1=\Omega_2=\Omega_3/2$.

\section{Model calculations and discussion}
\subsection{Adopted wavepackets}

There are many possibilities in choosing expansion coefficients $a_{\nu}$ ($\nu=0-3$)
of a wavepacket which satisfy Eq. (\ref{eq:G3}). 
Among them, we have studied in this paper, 
the four wavepackets A-D whose expansion coefficients are listed in Table 1.

\begin{center}
\begin{tabular}[b]{|c|c|c|c|c|c| }
\hline
wavepacket & $a_0$ & $a_1$ &  $ a_2$ & $ a_3 $  \\ 
\hline \hline
$\;\;\;$ A $\;\;\;$ &  $\;\;\;\; \frac{1}{2} \;\;\;\;$  &
  $\;\; \frac{1}{\sqrt{2}} \;\;$ 
& $\;\;\; 0 \;\;\;$ & $\;\;\; \frac{1}{2} \;\;\;$  \\
\hline
B   &  $\frac{1}{\sqrt{2}}$  &  0  & 0 & $\frac{1}{\sqrt{2}}$   \\
\hline
C&  $\frac{1}{\sqrt{2}} $ &  $\frac{1}{\sqrt{2}}$ & 0 &  0 \\
\hline
D  &  $\frac{1}{2}$ &  $\;\;\frac{1}{2} \;\;$ & 
$\;\;\frac{1}{2}\;\;$ & $\frac{1}{2}$ \\
\hline
\end{tabular}
\end{center}
{\it Table 1} Assumed expansion coefficients $a_{\nu}$ ($\nu=0$ to 3) for
four wavepackets A, B, C and D. 

\begin{figure}
\begin{center}
\includegraphics[keepaspectratio=true,width=120mm]{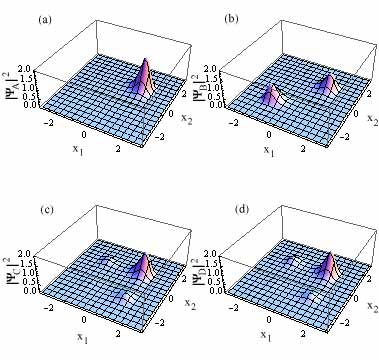}
\end{center}
\caption{
(Color online) 
Magnitudes $\vert \Psi(x_1,x_2) \vert^2$ of adopted four wavepackets of (a) A, (b) B, 
(c) C and (d) D for $g=0.0$ at $t=0.0$.
}
\label{fig3}
\end{figure}

Coefficients in adopted wavepackets A-D are chosen as follows:
A factorizable product state for $g=0.0$ is expressed by
\begin{eqnarray}
\Psi_{prod}&=& 
\Psi_R(x_1) \Psi_R(x_2), 
\label{eq:A1a}\\
&=& \frac{1}{2} \left[ \phi_0(x_1) \phi_0(x_2)+ \phi_0(x_1) \phi_1(x_2) 
+\phi_1(x_1) \phi_0(x_2)+\phi_1(x_1) \phi_1 (x_2)\right], \\
&=& \frac{1}{2} \left[ \Phi_0(x_1,x_2) +\Phi_3(x_1,x_2)  \right]
+ \frac{1}{\sqrt{2}} \Phi_1(x_1,x_2),
\label{eq:A1}
\end{eqnarray}
where magnitude of $\Psi_R(x_{\nu})$ $( =[\phi_0(x_{\nu})+\phi_1(x_{\nu})]/\sqrt{2} )$ 
localizes at the right well in the $x_{\nu}$ axis ($\nu=1, \:2$).
The wavepacket yielding initially the product state given by Eq. (\ref{eq:A1})
is described by the wavepacket A with $a_0=a_3=1/2$ and $a_1=1/\sqrt{2}$.

As a typical entangled state which cannot be expressed in a factorized form, 
we consider the state for $g=0.0$,
\begin{eqnarray}
\Psi_{ent}(x_1,x_2) &=& 
\frac{1}{\sqrt{2}} \left[ \phi_0(x_1) \phi_0(x_2)+\phi_1(x_1) \phi_1(x_2)  \right], \\
&=& \frac{1}{\sqrt{2}} \left[ \Phi_0(x_1,x_2) +\Phi_3(x_1,x_2)  \right].
\label{eq:B1}
\end{eqnarray}
The relevant wavepacket is expressed by the wavepacket B with $a_0=a_3=1/\sqrt{2}$.

The wavepacket C consists of the ground and first-excited states
with $a_0=a_1=1/\sqrt{2}$, which has been commonly adopted as a wavepacket. 
The wavepacket D includes four components with equal weights of $a_{\nu}=1/2$ for $\nu=0-3$.

Figures 3(a), 3(b), 3(c) and 3(d) show magnitudes $\vert \Psi(x_1,x_2) \vert^2$ 
of wavepackets A, B, C and D, respectively, for $g=0.0$ at $t=0.0$ generated 
by Eq. (\ref{eq:G1}) with expansion coefficients shown in Table 1.
The wavepacket A has a peak at the RR side in the $(x_1,x_2)$ space while
the wavepacket B has two peaks at RR and LL sides,
where $RR$ ($LL$) signifies the right (left) side in the $x_1$ axis and 
the right (left) side in $x_2$ axis.
Wavepackets C and D have similar profiles with main peaks at the RR side
at $t=0.0$ for $g=0.0$, but they are
quite different at $t \neq 0.0$ or for $g \neq 0.0$ 
(compare Figs. \ref{fig8} and \ref{fig9} with Figs. \ref{fig10} and \ref{fig11}, respectively). 
Wavepackets A, B, C and D which are initially localized in the $(x_1, x_2)$ space
are expected to be meaningful among conceivable wavepackets.

\subsection{Dynamics of $\Gamma(t)$ and $C(t)$}
We will study dynamics of $\Gamma(t)$ and $C(t)$ for wavepackets A, B, C and D,
which are separately described in subsections 1, 2, 3 and 4, respectively \cite{Note2}.

\begin{figure}
\begin{center}
\includegraphics[keepaspectratio=true,width=80mm]{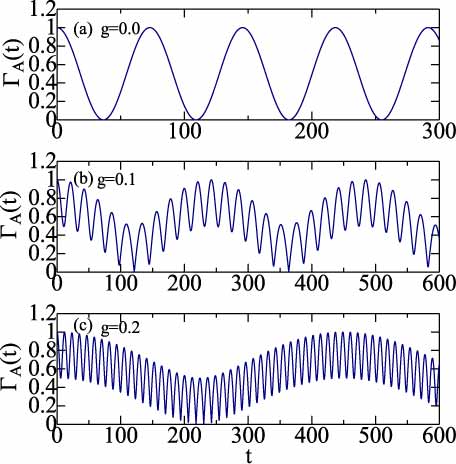}
\end{center}
\caption{
(Color online) 
Correlation function $\Gamma_A(t)$ for the wavepacket A 
with (a) $g=0.0$, (b) $g=0.1$ and (c) $g=0.2$.
}
\label{fig4}
\end{figure}

\begin{figure}
\begin{center}
\includegraphics[keepaspectratio=true,width=80mm]{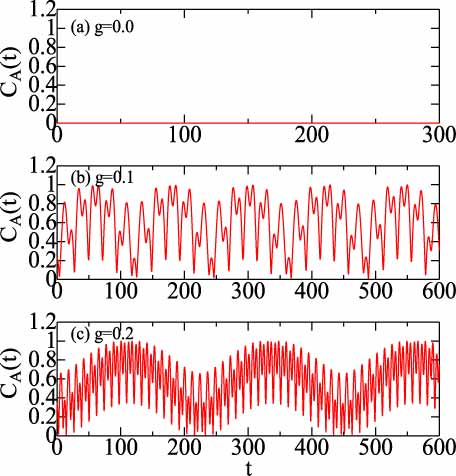}
\end{center}
\caption{
(Color online) 
Concurrence $C_A(t)$ for the wavepacket A 
with (a) $g=0.0$, (b) $g=0.1$ and (c) $g=0.2$, $C_A(t)$ being vanishing in (a).
}
\label{fig5}
\end{figure}
 
\subsubsection{Wavepacket A: $a_0=1/2$, $a_1=1/\sqrt{2}$, $a_2=0$ and $a_3=1/2$}
From Eq. (\ref{eq:G2}) and expansion coefficients in Table 1,
the correlation function of the wavepacket A is given by
\begin{eqnarray}
\Gamma_A &=& \vert \frac{1}{2} e^{-i\Omega_1 t}
+\frac{1}{4} (1+e^{-i\Omega_3 t})\vert.
\end{eqnarray}
Figure \ref{fig4}(a) shows the correlation function $\Gamma_A(t)$
calculated for $g=0.0$ which yields $\tau=36.40$. 
Figures \ref{fig4}(b) and \ref{fig4}(c) show $\Gamma_A(t)$ with $g=0.1$ and $0.2$, 
respectively, which oscillate more rapidly than that with $g=0.0$
in Fig. \ref{fig4}(a). However, the orthogonality times for $g=0.1$ and 0.2
are given by $\tau= 121.0$ and $218.8$, respectively, 
which are larger than that for $g=0.0$ (36.40).

From Eq. (\ref{eq:K4b}), the concurrence of the wavepacket A is given by
\begin{eqnarray}
C_A(t)^2 &=& \frac{1}{16} \;\vert 2\:e^{-2i \Omega_1 t}- 2 \cos 2 \theta 
\:e^{-i \Omega_3 t}- \sin 2 \theta (1-e^{-2i \Omega_3 t}) \vert^2,
\end{eqnarray}
which reduces to
\begin{eqnarray}
C_A(0)^2 &=& \frac{1}{4} \;(1- \cos 2 \theta)^2. 
\end{eqnarray}
Figure \ref{fig5}(a) shows that $C_A(t)$ for $g=0.0$ is vanishing independently of time.
We note in Figs. \ref{fig5}(b) and \ref{fig5}(c) that
when the coupling is introduced, 
$C_A(t)$ with initial values of $C_A(0)=0.223$ and 0.342
for $g=0.1$ and $0.2$, respectively, show complex time dependence
which arises from a superposition of multiple contributions with
frequencies of $\Omega_1$, $\Omega_3$, $\Omega_3-\Omega_1$ and $\Omega_3-2 \Omega_1$. 

The temporal average of $\langle C_A(t)^2 \rangle$ may be analytically calculated as
\begin{eqnarray}
C_{av}^2 = \langle C_A(t)^2 \rangle &=&
\frac{1}{8} (4 - \sin^2 2 \theta) \hspace{1cm} \mbox{for $g > 0.0$}, \nonumber \\
&=& 0 \hspace{3cm}\mbox{for $g=0.0$}.
\end{eqnarray}
Note that $C^2_{av}$ has the discontinuity at $g=0.0$
where $\Omega_3-2 \Omega_1 = 0$ (Fig. 2) \cite{Note1}.
We obtain $C_{av}=0.0$, 0.707, 0.643 and 0.622 
for $g=0$, $0_+$, 0.1 and 0.2, respectively, 
where $0_+ = \lim_{\epsilon \rightarrow 0} 0+\epsilon$.

\subsubsection{Wavepacket B: $a_0=1/\sqrt{2}$, $a_1=a_2=0.0$ and $a_3=1/\sqrt{2}$}
\begin{figure}
\begin{center}
\includegraphics[keepaspectratio=true,width=80mm]{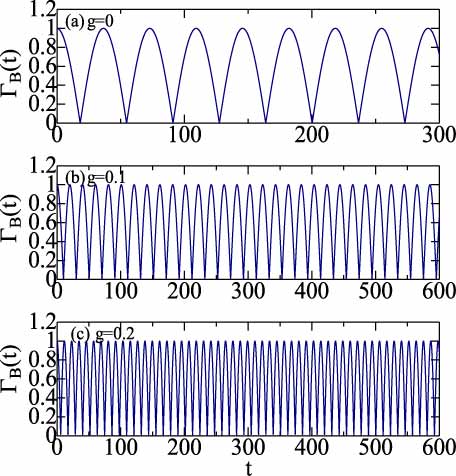}
\end{center}
\caption{
(Color online) 
Correlation function $\Gamma_B(t)$ for the wavepacket B 
with (a) $g=0.0$, (b) $g=0.1$ and (c) $g=0.2$.
}
\label{fig6}
\end{figure}

The correlation function of the wavepacket B is given by
\begin{eqnarray}
\Gamma_B(t) &=& \frac{1}{2}(1+e^{-i \Omega_3 t}),
\end{eqnarray}
which leads to $\tau=\pi/\Omega_3$.
Figures \ref{fig6}(a),  \ref{fig6}(b) and \ref{fig6}(c) show
$\Gamma_B(t)$ for $g=0.0$, 0.1 and 0.2, respectively, for which
the orthogonality times are $\tau=18.2$, 10.1 and 5.75.

\begin{figure}
\begin{center}
\includegraphics[keepaspectratio=true,width=80mm]{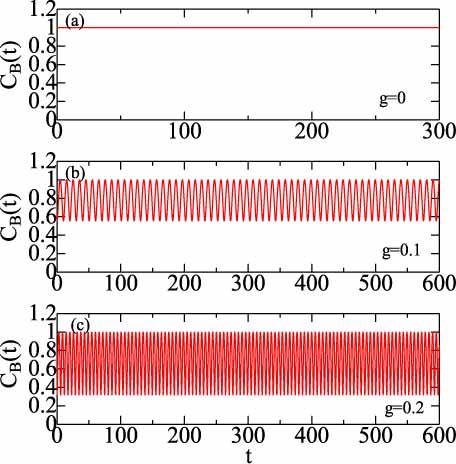}
\end{center}
\caption{
(Color online) 
Concurrence $C_B(t)$ for the wavepacket B 
with (a) $g=0.0$, (b) $g=0.1$ and (c) $g=0.2$.
}
\label{fig7}
\end{figure}

\begin{figure}
\begin{center}
\includegraphics[keepaspectratio=true,width=80mm]{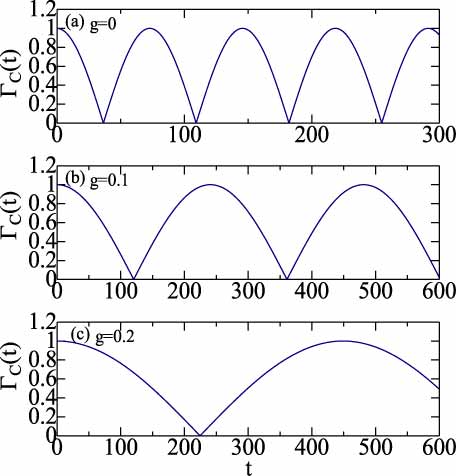}
\end{center}
\caption{
(Color online) 
Correlation function $\Gamma_C(t)$ for the wavepacket C
with (a) $g=0.0$, (b) $g=0.1$ and (c) $g=0.2$.
}
\label{fig8}
\end{figure}

\begin{figure}
\begin{center}
\includegraphics[keepaspectratio=true,width=80mm]{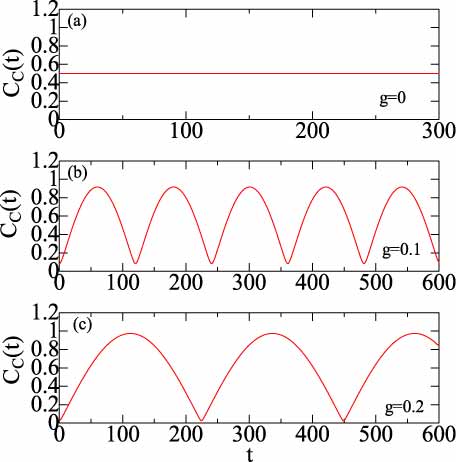}
\end{center}
\caption{
(Color online) 
Concurrence $C_C(t)$ for the wavepacket C 
with (a) $g=0.0$, (b) $g=0.1$ and (c) $g=0.2$.
}
\label{fig9}
\end{figure}

The concurrence of the wavepacket B is expressed by
\begin{eqnarray}
C_B(t)^2 &=& \frac{1}{4}\vert \sin 2 \theta (1-e^{-2 i \Omega_3 t})
+2 \cos 2 \theta \:e^{-i \Omega_3 t} \vert^2,
\end{eqnarray}
with
\begin{eqnarray}
C_B(0)^2 &=& \cos^2 2 \theta,
\label{eq:KB1}
\end{eqnarray}
from which we obtain $C_B(0)=1.0$, 0.554 and 0.316 for $g=0.0$, 0.1 and 0.2, respectively.
Calculated $C_B(t)$ for $g=0.0$, 0.1 and 0.2 are plotted in
Figs. \ref{fig7}(a), \ref{fig7}(b) and \ref{fig7}(c), respectively.
$C_B(t)$ for $g=0.0$ is unity independently of time. When the coupling $g$ is
introduced, $C_B(t)$ becomes time dependent, showing rapid oscillations
as shown in Figs. \ref{fig7}(b) and \ref{fig7}(c).

The temporal average of $\langle C_B(t)^2 \rangle$ is given by
\begin{eqnarray}
C_{av}^2 &=& \langle C_B(t)^2 \rangle =1 - \frac{1}{2}\sin^2 2 \theta,
\label{eq:KB2}
\end{eqnarray}
which leads to $C_{av}=1.0$, 0.808 and 0.742 for $g=0.0$, 0.1 and 0.2, respectively.

\subsubsection{Wavepacket C: $a_0=a_1=1/\sqrt{2}$ and $a_2=a_3=0.0$}
The correlation function of the wavepacket C is given by
\begin{eqnarray}
\Gamma_C(t) &=& \frac{1}{2}(1+e^{-i \Omega_1 t}),
\end{eqnarray}
leading to $\tau=\pi/\Omega_1$.
Figures \ref{fig8}(a), \ref{fig8}(b) and \ref{fig8}(c) show
$\Gamma_C(t)$ for $g=0.0$, 0.1 and 0.2, respectively, from which
the orthogonality time is given by $\tau=36.40$, 120.3 and 224.5. 

The concurrence of the wavepacket C is expressed by
\begin{eqnarray}
C_C(t)^2 &=& \frac{1}{4} \;\vert e^{-2 i \Omega_1 t}- \sin{2 \theta} \vert^2,
\end{eqnarray}
which reduces to
\begin{eqnarray}
C_C(0)^2 &=& \frac{1}{4} (1- \sin 2 \theta)^2.
\end{eqnarray}
We obtain $C_C(0)=0.5$, 0.0839 and 0.0256 for $g=0.0$, 0.1 and 0.2, respectively. 
Figure \ref{fig9}(a) shows the time-independent $C_C(t)=0.5$ for $g=0.0$.
For $g=0.1$ and $0.2$, $C_C(t)$ show oscillations as shown 
in Figs. \ref{fig9}(b) and \ref{fig9}(b).

The temporal average is given by
\begin{eqnarray}
C_{av}^2 &=& \langle C_C(t)^2
= \frac{1}{4} (1+ \sin^2 2 \theta),
\end{eqnarray}
which yields $C_{av}=0.5$, 0.651 and 0.689 for $g=0.0$, 0.1 and 0.2, respectively.

\begin{figure}
\begin{center}
\includegraphics[keepaspectratio=true,width=80mm]{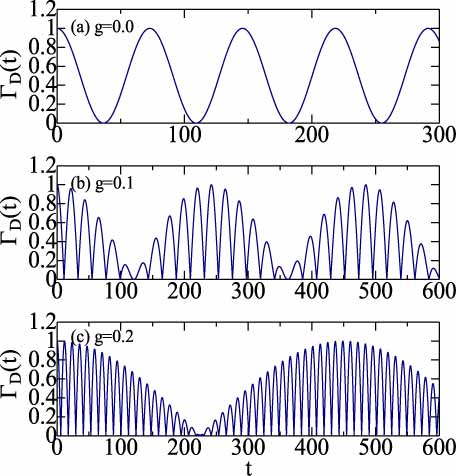}
\end{center}
\caption{
(Color online) 
Correlation function $\Gamma_D(t)$ for the wavepacket D
with (a) $g=0.0$, (b) $g=0.1$ and (c) $g=0.2$.
}
\label{fig10}
\end{figure}

\begin{figure}
\begin{center}
\includegraphics[keepaspectratio=true,width=80mm]{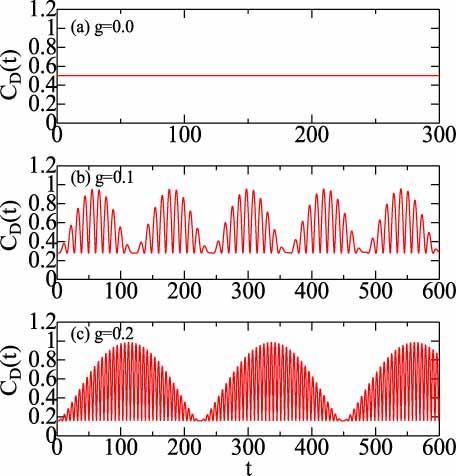}
\end{center}
\caption{
(Color online) 
Concurrence $C_D(t)$ for the wavepacket D 
with (a) $g=0.0$, (b) $g=0.1$ and (c) $g=0.2$.
}
\label{fig11}
\end{figure}

\subsubsection{Wavepacket D: $a_0=a_1=a_2=a_3=1/2$}
The correlation function of the wavepacket D is expressed by
\begin{eqnarray}
\Gamma_D(t) &=& \frac{1}{4} \vert 1+e^{-i\Omega_1 t}
+e^{-i\Omega_2 t}+e^{-i\Omega_3 t} \vert, \\
&=& \frac{1}{4} \vert (1+e^{-i\Omega_1 t}) (1+e^{-i\Omega_2 t}) \vert,
\end{eqnarray}
where Eq. (\ref{eq:H8}) is employed.
For $g=0.0$, $\Gamma_D(t)$ shows a simple sinusoidal oscillation
because $\Omega_1=\Omega_2=\Omega_3/2$ [Fig. \ref{fig10}(a)].
For $g \neq 0.0$, however, $\Gamma_D(t)$ exhibits a rather complex oscillation
as shown in Figs. \ref{fig10}(b) and \ref{fig10}(c), where
$\Gamma_D(t)$ vanishes at $t=36.40 \:(2k+1)$, $11.02 \:(2k+1)$ and $5.903 \:(2k+1)$
for $g=0.0$, 0.1 and 0.2, respectively, with $k=0,1,2 \cdots $. We obtain
the orthogonality time expressed by $\tau = \pi/\Omega_2$, which leads to
$\tau=36.40$, 11.02 and 5.903 for $g=0.0$, 0.1 and 0.2, respectively.

\begin{figure}
\begin{center}
\includegraphics[keepaspectratio=true,width=120mm]{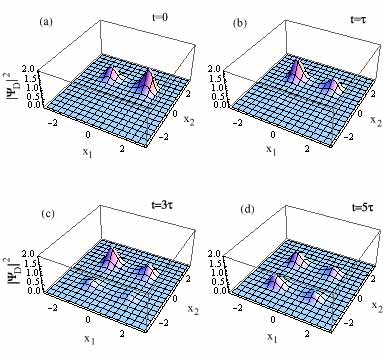}
\end{center}
\caption{
(Color online) 
Magnitudes of $\vert \Psi_D(x_1, x_2,t) \vert^2$ of the wavepacket D with $g=0.1$ 
at (a) $t=0$, (b) $t=\tau$, (c) $t=3 \tau$ and (d) $t=5 \tau$,
$\Psi_D(x_1, x_2, t)$ at $t= \tau$, $3 \tau$ and $5 \tau$ being orthogonal to $\Psi_D(x_1, x_2, 0)$.
}
\label{fig12}
\end{figure}

The concurrence of the wavepacket D is given by
\begin{eqnarray}
C_D(t)^2 &=& \frac{1}{16} \; \vert \sin 2 \theta (1-e^{-2i \Omega_3 t})
+2 \cos{2 \theta} \:e^{-i \Omega_3 t} -e^{-2i \Omega_1 t}+e^{-2 i \Omega_2 t} \vert^2,
\end{eqnarray}
with
\begin{eqnarray}
C_D(0)^2 &=& \frac{1}{4} \cos^2 2 \theta,
\end{eqnarray}
yielding  $C_D(0)=0.5$, 0.277 and 0.158 for $g=0.0$, 0.1 and 0.2, respectively.
Although $C_D(t)$ for $g=0.0$ is 0.5 independently of $t$ in Fig. \ref{fig11}(a),
$C_D(t)$ for $g=0.1$ and 0.2 show complicated time dependence
in Figs. \ref{fig11}(b) and \ref{fig11}(c), respectively.

The averaged concurrence is given by 
\begin{eqnarray}
C_{av}^2 = \langle C_D(t)^2 \rangle 
&=& \frac{1}{8}(3-\sin^2 2\theta) \hspace{1cm} \mbox{for $g > 0.0$}, \\
&=& \frac{1}{4} \hspace{3cm}\mbox{for $g=0.0$},
\end{eqnarray}
where a discontinuity arises from the relation: 
$\Omega_1-\Omega_2 \rightarrow 0$ as $g \rightarrow 0.0$ \cite{Note1}.
We obtain $C_{av}=0.5$, 0.612, 0.537 and 0.512 for $g=0$, $0_+$, 0.1 and 0.2, respectively.

Before closing the subsection of Sec. III B,
it is worthwhile to make a closer look to the dynamical properties of wavefunctions.
There is one kind of wavepackets which is orthogonal to the initial 
wavepacket A, B or C: 
for example, $\vert \Psi_A(x_1,x_2, \tau (2k+1)) \vert^2$ with $k=0,1,2,\cdots$ 
has a peak at the LL side while $\vert \Psi_A(x_1,x_2, 0) \vert^2$ at the RR side.
It is, however, not the case for the wavepacket D.
Magnitudes of wavefunctions 
$\vert \Psi_D(x_1, x_2,t) \vert^2$ for $g=0.1$ at $t=0.0$, $\tau$, $3 \tau$ and $5 \tau$
($\tau=11.02$) are plotted in Figs. \ref{fig12}(a)-\ref{fig12}(d), respectively,
where all wavefunctions in Figs. \ref{fig12}(b), \ref{fig12}(c) and \ref{fig12}(d)
are orthogonal to that in Fig. \ref{fig12}(a).

\subsection{The $g$ dependence of $\tau$ and $\tau_{min}$}
\begin{figure}
\begin{center}
\includegraphics[keepaspectratio=true,width=120mm]{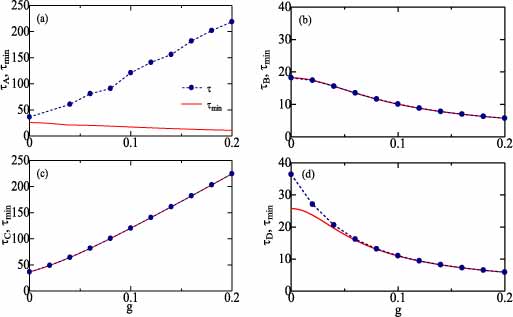}
\end{center}
\caption{
(Color online) 
The $g$ dependence of the orthogonality time $\tau$ (dashed curves)
and minimum values $\tau_{min}$ (solid curves)
for (a) wavepacket A, (b) B, (c) C and (d) D.
}
\label{fig13}
\end{figure}

Although calculations in the preceding subsection Sec. III B have been reported only 
for $g=0.0$, 0.1 and 0.2, we may repeat calculations of 
the orthogonality time $\tau$ by changing $g$ for the four wavepackets. 
Calculated $\tau$ is plotted as a function of $g$ by dashed curves
in Figs. \ref{fig13}(a)-\ref{fig13}(d). 
Obtained $\tau$ for the four wavepackets is expressed in the second column of Table 2,
whose third and fourth columns show $C^2_{av}$ and $C(0)^2$, respectively,
$E_{\nu}$ ($\nu=0 - 3$) and $\theta$ being $g$ dependent 
[Eqs. (\ref{eq:H6a})-(\ref{eq:H6b}), (\ref{eq:H7c})].
Figures \ref{fig13}(a)-\ref{fig13}(d) show that
with increasing $g$, $\tau_A$ and $\tau_C$ are increased, while $\tau_B$ and $\tau_D$
are decreased. This is because with increasing $g$, a gap of $E_1-E_0$ is decreased
whereas $E_3-E_0$ and $E_2-E_0$ are increased (Fig. 2).

\begin{center}
\begin{tabular}[t]{|c|c|c|c|c| }
\hline
wavepacket & $\;\;\; \tau \;\;\;$ & $C^2_{av}$ &  $ C(0)^2 $   \\ 
\hline \hline
$\;\;\;$ A $\;\;\;$ &  $\;\;\;\; \simeq  \frac{\pi \hbar}{(E_1-E_0)} \;\;\;\;$  &
  $\;\; \frac{1}{8}(4- \sin^2 2 \theta)-\frac{1}{2}\:\delta(g) \;\;$ 
& $\;\;\; \frac{1}{4}(1 - \cos 2 \theta)^2 \;\;\;$   \\
\hline
B   &  $\frac{\pi \hbar}{(E_3-E_0)}$  &  $1- \frac{1}{2} \sin^2 2 \theta$  
& $\cos^2 2 \theta $ \\
\hline
C&  $ \;\;\;\frac{\pi \hbar}{(E_1-E_0)} \;\;\; $ &  $\frac{1}{4}(1+\sin^2 2 \theta)$ 
& $\frac{1}{4}(1- \sin 2 \theta)^2$  \\
\hline
D  &  $ \frac{\pi \hbar}{(E_2-E_0)}$ 
&  $\frac{1}{8}(3- \sin^2 2 \theta) -\frac{1}{8} \:\delta(g)$ 
& $\frac{1}{4} \cos^2 2 \theta $  \\
\hline
\end{tabular}
\end{center}
{\it Table 2} Calculated $\tau$, $C^2_{av}$ and $C(0)^2$ for four wavepackets A, B, C and D. 


\begin{figure}
\begin{center}
\includegraphics[keepaspectratio=true,width=120mm]{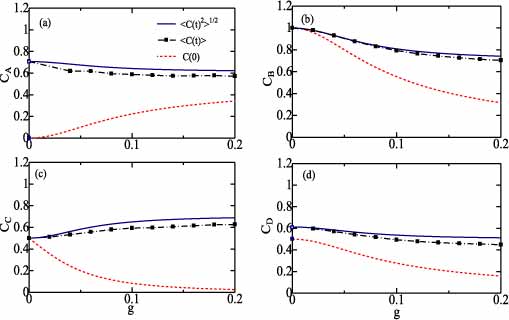}
\end{center}
\caption{
(Color online) 
The $g$ dependence of averaged concurrences, 
$C_{av}$ [$=\sqrt{\langle C(t)^2\rangle}$] (solid curves) and
$\langle C(t) \rangle$ (chain curves),
and of the initial concurrence $C(0)$ (dashed curves) 
for (a) wavepacket A, (b) B, (c) C and (d) D.
}
\label{fig14}
\end{figure}

We may evaluate the minimum orthogonality time $\tau_{min}$ of our DW model,
calculating the expectation energy $E$ and its root-mean-square value $\Delta E$ in
Eq. (\ref{eq:J1}), which are expressed by
\begin{eqnarray}
E 
&=& \sum_{\nu}\: \vert a_{\nu}\vert^2 \:(E_{\nu}-E_0)
= \vert a_1 \vert^2 \Omega_1+ \vert a_2 \vert^2 \Omega_2+ \vert a_3 \vert^2 \Omega_3, 
\label{eq:J2}\\
(\Delta E)^2 
&=& \sum_{\nu}\: \vert a_{\nu}\vert^2 \:(E_{\nu}-E_0)^2-E^2
= \vert a_1 \vert^2 \Omega_1^2+ \vert a_2 \vert^2 \Omega_2^2+ \vert a_3 \vert^2 \Omega_3^2-E^2.
\label{eq:J3}
\end{eqnarray}
Solid curves in Figs. \ref{fig13}(a)-\ref{fig13}(d) express
$\tau_{min}$ calculated by Eqs. (\ref{eq:J1}), (\ref{eq:J2}) and  (\ref{eq:J3}) 
as a function of $g$ for four wavepackets A-D. 
With increasing $g$, $\tau_{min}$ increases for the wavepacket C, whereas
those for wavepackets A, B and D decrease.
The ratio of $\tau/\tau_{min}$ is unity for wavepackets B and C 
in Figs. \ref{fig13}(b) and \ref{fig13}(c). However, we obtain
$\tau/\tau_{min} = 1.414$ and 1.0 for $g=0.0$ and $g \gtrsim 0.06$, respectively, 
for the wavepacket D in Fig. \ref{fig13}(d). Furthermore,
this ratio more apparently exceeds unity for the wavepacket A in Fig. \ref{fig13}(a) where
$\tau/\tau_{min}$ $=1.414$, 7.00 and 20.0 for $g=0.0$, 0.1 and 0.2, respectively. 
This is accounted for by the fact that
with increasing $g$, $\tau$ is increased because of a narrowed energy gap of $E_1-E_0$ 
while $\tau_{min}$ is decreased by a high-energy contribution 
of $E_3-E_0$ to $E$ (Fig. \ref{fig2}) . 

\subsection{$g$ dependence of $C_{av}$ and $C(0)$}
We may calculate $C_{av}$ and $C(0)$ of the four wavepackets
as a function of $g$, whose results are plotted in Figs. \ref{fig14}(a)-\ref{fig14}(d).
In the wavepacket A, $C_{av}$ has a discontinuity at $g=0.0$ as mentioned before:
$C_{av}=0.0$ and 0.707 at $g=0$ and $0_+$, respectively.
When $g$ is introduced, $C(0)$ is increased from zero while $C_{av}$ is decreased from $0.707$
[Fig. \ref{fig14}(a)].
In the wavepacket B, both $C_{av}$ and $C(0)$ are gradually decreased
with increasing $g$ [Fig. \ref{fig14}(b)].
On the contrary, in the wavepacket C, $C_{av}$ is increased from 0.5
but $C(0)$ is decreased form 0.5 when $g$ is introduced [Fig. \ref{fig14}(c)].
In the wavepacket D, $C_{av}$ has a discontinuity at $g=0.0$:
$C_{av}=0.5$ and 0.612 at $g=0$ and $0_+$, respectively, and both $C_{av}$ and $C(0)$
are decreased with increasing $g$ [Fig. \ref{fig14}(d)].
 
Chain curves in Figs. \ref{fig14}(a)-\ref{fig14}(d) show $\langle C(t) \rangle$
for the four wavepackets, which are nearly in agreement
with $C_{av}$ ($=\sqrt{C(t)^2}$) plotted by solid curves.

\begin{figure}
\begin{center}
\includegraphics[keepaspectratio=true,width=80mm]{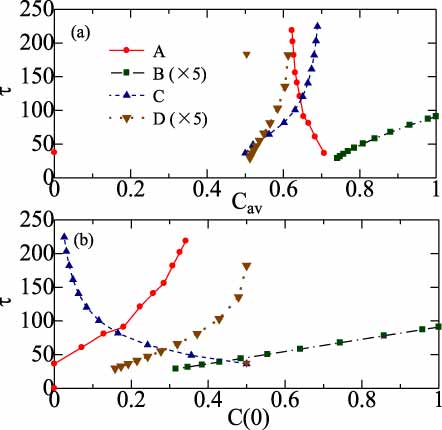}
\end{center}
\caption{
(Color online) 
(a) $C_{av}$ vs. $\tau$ and (b) $C(0)$ vs. $\tau$
for wavepackets A (circles), B (squares), 
C (triangles) and D (inverted triangles),
$\tau$ for wavepackets B and D being multiplied by a factor of five.
In (a), a point at $(C_{av}, \tau)=(0.0, 36.40)$ for the wavepacket A
is separated from that at $(0.706, 36.40)$
because of the discontinuity of $C_{av}$ at $g=0.0$ (Fig. \ref{fig14}).
Similarly, a point at $(C_{av},\tau)=(0.5, 36.40)$ for the wavepacket D is separated 
from that at $(0.6124, 36.40)$.
}
\label{fig15}
\end{figure}

\subsection{The dependence of $\tau$ on $C_{av}$ and $C(0)$}
Comparing Figs. \ref{fig13}(a)-\ref{fig13}(d) and Figs. \ref{fig14}(a)-\ref{fig14}(d), respectively,
we may examine the relation between the orthogonality time ($\tau$) and the concurrence
($C_{av}$ or $C(0)$).
Figures \ref{fig15}(a) and \ref{fig15}(b) show the $C_{av}$ vs. $\tau$ plot
and the $C(0)$ vs. $\tau$ plot, respectively.
For wavepackets B and D (squares and inverted triangles), $\tau$ 
is increased with increasing $C_{av}$ or $C(0)$.
We note, however, that $\tau$ of the wavepacket A (circles) is increased 
with increasing $C(0)$ but with decreasing $C_{av}$.
On the contrary, $\tau$ of the wavepacket C (triangles) is increased 
with increasing $C_{av}$ but with decreasing $C(0)$.
Figures \ref{fig15}(a) and \ref{fig15}(b) imply that the effect of $C_{av}$
on $\tau$ is generally different from that of $C(0)$ and
that when the concurrence is increased,
the orthogonality time may be increased or decreased, depending on the adopted wavepacket.
This is in contrast with Refs. \cite{Giovannetti03,Batle05,Curilef06}
but in agreement with Refs. \cite{Giovannetti03b,Borras06,Zander13}.

\section{Concluding remark}
Batle {\it et al.} \cite{Batle05} and Curilef {\it et al.} \cite{Curilef06}
studied two uncoupled qubits with eigenvalues
\begin{eqnarray}
E_0 = 0, \;\;\; E_1=E_2=\epsilon_1,
\;\;\; E_3=2 \epsilon_1,   
\label{eq:J6}
\end{eqnarray}
where $\epsilon_1$ stands for an energy of a free qubit.
For wavepackets with $\vert a_0 \vert^2= \vert a_3 \vert^3 \neq 0.0$,
the ratio of $\tau/\tau_{min}$ is shown to be 
unity for a maximally entangled state
and $\sqrt{2}$ for a separate state \cite{Batle05,Curilef06}.
In our wavepackets A, B and D with $a_3=a_0 \neq 0.0$ for $g=0.0$, 
the ratio is $\tau/\tau_{min}=1.0$
in the wavepacket B while it is $\sqrt{2}$ in wavepackets A and D,
which are consistent with results of Refs. \cite{Batle05,Curilef06}.
However, in the wavepacket C with $a_3=0.0$, which corresponds to 
the {\it singular} case after Chau \cite{Chau10}, we obtain $\tau/\tau_{min}=1.0$ for $g=0.0$ 
although it is not a maximally entangled state ($C=0.5$),
in agreement with Ref. \cite{Chau10}.

Zander {\it et al.} \cite{Zander13} adopted two interacting qubits given by the Hamiltonian
\begin{eqnarray}
H &=& \hbar \omega_0 [2 I^{(1)} \otimes I^{(2)}
-\sigma_x^{(1)} \otimes I^{(2)}- I^{(1)} \otimes\sigma_x^{(2)}]
+ \hbar \omega [I^{(1)} \otimes I^{(2)}-\sigma_x^{(1)} \otimes \sigma_x^{(2)}],
\label{eq:J4}
\end{eqnarray}
where $\omega_0$ expresses the energy of free qubits, $\omega$ stands for the interaction, 
$I$ is the identity matrix and $\sigma_x$ is the $x$-Pauli matrix.
Eigenvalues of the Hamiltonian are \cite{Zander13}
\begin{eqnarray}
E_0 = 0, \;\;\; E_1=E_2=2 \hbar(\omega+\omega_0),
 \;\;\; E_3=4 \hbar \omega_0.   
\label{eq:J5} 
\end{eqnarray}
Ref. \cite{Zander13} studied effects of entanglement on the evolution speed 
in interacting two qubits, evaluating the linear entropy
mainly for the three cases of $\omega=\omega_0$, $\omega_0=0$ and $\omega=3 \omega_0$ 
with arbitrary expansion coefficients $\{ a_{\nu} \}$ for wavepackets.
The study of Ref. \cite{Zander13} is complementary to ours 
in which calculations have been made 
for an arbitrary interaction $g$ with four sets of expansion coefficients $\{ a_{\nu} \}$ 
for wavepackets A-D.
It was claimed in Ref. \cite{Zander13} that with the exception of some {\it special} cases, states with a small 
initial entanglement tend to evolve in the fastest way in coupled qubits.
This is not inconsistent with our result of the wavepacket A showing that $\tau/\tau_{min}=7.00$ 
and $20.0$ for $C(0)=0.223$ and 0.342, respectively.
However, we obtain $\tau/\tau_{min}=1.0$ almost independently of $C(0)$ 
in wavepackets B, C, and D (Figs. \ref{fig13} and \ref{fig14}), which might correspond to
special cases after Ref. \cite{Zander13}. 
Refs. \cite{Giovannetti03} and \cite{Zander13} explained that for $\tau$ to reach the bound,
it is necessary to have either an initial entangled state, or an interaction term
capable of creating entanglement.
This seems not to be applicable to the wavepacket A for which 
$\tau/\tau_{min}=20.0 \gg 1.0$ even if $C(0)=0.342$ for $g=0.2$ 
(Figs. \ref{fig13} and \ref{fig14}).
This disagreement might arise from a difference in models adopted in Ref. \cite{Zander13}
and the present study: the interaction dependence of eigenvalues
in Eq. (\ref{eq:J5}) are different from that in Eqs. (\ref{eq:H6a})-(\ref{eq:H6b}).

In the simple case, we may obtain an analytical expression
for $\tau$ expressed in terms of $C_{av}$ and/or $C(0)$.
Indeed, for the wavepacket B, a calculation with 
Eqs. (\ref{eq:H7c}), (\ref{eq:KB1})and (\ref{eq:KB2}) leads to
\begin{eqnarray}
\tau_B &=& \tau_{min}= \left( \frac{\pi}{2 \delta} \right)\:\sqrt{2 C_{av}^2-1} 
= \left( \frac{\pi}{2 \delta} \right) C(0),
\label{eq:KB3}
\end{eqnarray}
which is numerically confirmed in Fig. \ref{fig15}.
Unfortunately it is impossible to obtain analytical results for wavepackets A, C and D.

In summary, we have studied dynamical properties of four wavepackets A, B, C and D (Table 1),
by using an exactly solvable coupled DW system described by Razavy's potential \cite{Razavy80}.
Our model calculations yield the followings:

\noindent
(1) The correlation function $\Gamma(t)$ and concurrence $C(t)$ in interacting two qubits  
show complicate and peculiar time dependence (Figs. \ref{fig4}-\ref{fig11}),
 
\noindent 
(2) The quantum evolution speed measured by $\tau^{-1}$ is not necessarily
increased by an introduced interaction $g$:
{\it e.g. } it is decreased in wavepackets A and C (Fig. \ref{fig13}), 

\noindent
(3) The concurrence, $C_{av}$ or $C(0)$, may be decreased by 
an increased interaction (Fig. \ref{fig14}), 

\noindent
(4) The relation between $C(0)$ and $\tau$ is generally not the same as
that between $C_{av}$ and $\tau$, and the evolution speed may be increased or decreased 
with the increased concurrence ($C_{av}$ or $C(0)$), 
depending on a wavepacket (Fig. \ref{fig15}),
and

\noindent
(5) $\tau$ may not reach its minimum value $\tau_{min}$
even when the entanglement is present in coupled DW systems.

\noindent
Items (4) and (5) are in contrast with the non-interacting case
where $\tau$ is decreased with increasing $C(0)$ and the ratio $\tau/\tau_{min}$
approaches unity in entangled state
\cite{Giovannetti03,Giovannetti03b,Batle05,Borras06,Curilef06}.
Items (4) and (5) suggest that the relation between the evolution speed and the entanglement
in coupled qubits is not definite 
in contrast to that in uncoupled case
\cite{Giovannetti03,Giovannetti03b,Batle05,Borras06,Curilef06}.
In the present study, we do not take into account environmental effects which are expected
to play important roles in real DW systems. An inclusion of dissipative effects is left as
our future subject.

\begin{acknowledgments}
This work is partly supported by a Grant-in-Aid for Scientific Research from 
Ministry of Education, Culture, Sports, Science and Technology of Japan.  
\end{acknowledgments}



\begin{thebibliography}{99}
\bibitem{Ref1}
M. J. Storcz and F. K. Wilhelm,
Phys. Rev. A {\bf 67} (2003) 042319.

%
\bibitem{Margolus98}N. Margolus and L. B. Levitin,
Physica D {\bf 120} (1998) 188. 

\bibitem{Pfeifer93}L. Mandelstam and I. G. Tamm, 
J. Phys. USSR {\bf 9} (1945) 249;
%
K. Bhattacharyya, 
J. Phys. A {\bf 16} (1983) 2993;
%
P. Pfeifer,
Phys. Rev. Lett. {\bf 70} (1993) 3365.
%
%
\bibitem{Giovannetti03}V. Giovannetti, S. Lloyd, and L. Maccone,
Europhys. Lett. {\bf 62} (2003) 615.

\bibitem{Giovannetti03b}V. Giovannetti, S. Lloyd, and L. Maccone,
Phys. Rev. A {\bf 67} (2003) 052109.

\bibitem{Batle05}J. Batle, M. Casas, A. Plastino, and A. R. Plastino,
Phys. Rev. A {\bf 72} (2005) 032337.

\bibitem{Curilef06}S. Curilef, C. Zander and A. R. Plastino,
Eur. J. Phys. {\bf 27} (2006) 1193.

\bibitem{Borras06}A. Borr\'{a}s, M. Casas, A. R. Plastino, and A. Plastino,
Phys. Rev. A {\bf 74} (2006) 022326.

\bibitem{Chau10}H. F. Chau,
Phys. Rev. A {\bf 82} (2010) 056301.

\bibitem{Zander13}C. Zander, A. Borras, A. R. Plastino, A. Plastino, and M. Casas,
J. Phys. A {\bf 46}, 095302 (2013).

\bibitem{Tannor07}D. J. Tannor,
{\it Introduction to quantum mechanics: A time-dependent perspective}
(Univ. Sci. Books, Sausalito, California, 2007).

\bibitem{Razavy80}M. Razavy,
Am. J. Phys. {\bf 48} (1980) 285.

\bibitem{Finkel99}F. Finkel,
F. Finkel, A. Gonzalez-Lopez and M. A. Rodriguez,
J. Phys. A {\bf 32} (1999) 6821.

\bibitem{Bagchi03}B. Bagchi and A. Ganguly,
J. Phys. A {\bf 36} (2003) L161.

\bibitem{Note2}The energy and time are measured in units of
$\vert V(0) \vert/2$ and $2 \hbar/\vert V(0) \vert$, respectively, in this study.

\bibitem{Wootters01}W. K. Wootters,
Quan. Inf. Comp. {\bf 1} (2001) 27.

\bibitem{Note1}In the wavepacket A, one term in $\langle C_A(t)^2 \rangle$ leads to
$(1/t_f) \int_0^{t_f} \cos(\Omega_3-2 \Omega_1)t \:dt
=[\sin(\Omega_3-2 \Omega_1) t_f]/[(\Omega_3-2 \Omega_1)t_f]
\propto \delta(\Omega_3-2 \Omega_1)$ which yields the delta-function contribution at
$g=0$ where $\Omega_3-2 \Omega_1=0$.
The situation is similar to the wavepacket D in which $\Omega_2-\Omega_1 = 0$
at $g = 0$.

\end{thebibliography}
\end{document}